# Tunnel-injected sub-260 nm ultraviolet light emitting diodes

Yuewei Zhang,[1,a)] Sriram Krishnamoorthy,[1] Fatih Akyol,[1] Sanyam Bajaj,[1] Andrew A. Allerman,[2] Michael W. Moseley,[2] Andrew M. Armstrong,[2] and Siddharth Rajan[1,3,a)]

[1] Department of Electrical and Computer Engineering, The Ohio State University, Columbus, Ohio, 43210, USA

[2] Sandia National Laboratories, Albuquerque, New Mexico 87185, USA

[3] Department of Materials Science and Engineering, The Ohio State University, Columbus, Ohio, 43210, USA

**Abstract:** We report on tunnel-injected deep ultraviolet light emitting diodes (UV LEDs) configured with a polarization engineered $Al_{0.75}Ga_{0.25}N/ In_{0.25}Ga_{0.75}N$ tunnel junction structure. Tunnel-injected UV LED structure enables n-type contacts for both bottom and top contact layers. However, achieving Ohmic contact to wide bandgap n-AlGaN layers is challenging and typically requires high temperature contact metal annealing. In this work, we adopted a compositionally graded top contact layer for non-alloyed metal contact, and obtained a low contact resistance of $\rho_c=4.8\times10^{-5}$ $\Omega$ cm$^2$ on n-$Al_{0.75}Ga_{0.25}N$. We also observed a significant reduction in the forward operation voltage from 30.9 V to 19.2 V at 1 kA/cm$^2$ by increasing the Mg doping concentration from $6.2\times10^{18}$ cm$^{-3}$ to $1.5\times10^{19}$ cm$^{-3}$. Non-equilibrium hole injection into wide bandgap $Al_{0.75}Ga_{0.25}N$ with $E_g$>5.2 eV was confirmed by light emission at 257 nm. This work demonstrates the feasibility of tunneling hole injection into deep UV LEDs, and provides a novel structural design towards high power deep-UV emitters.

---

a) Authors to whom correspondence should be addressed.
   Electronic mail: zhang.3789@osu.edu, rajan@ece.osu.edu

III-Nitride ultraviolet light emitting diodes (UV LEDs) are promising in various applications including sterilization, water purification and medical sensing.[1] Research efforts over the past decade have led to the demonstration of UV light emission over a wide wavelength range from 400 nm to 210 nm.[2-5] Considerable improvements in substrate and active region quality have been achieved by optimizing the growth techniques, resulting in high radiative efficiency (~ 80%).[2] However, current UV LEDs exhibit significantly lower wall-plug efficiency as compared to their blue LED counterparts.[1-3,6,7]

The limitation has been attributed to the high p-type contact resistance and low conductivity of p-type AlGaN layers. Poor direct p-type contact to AlGaN layers makes it necessary to use a thick p-GaN cap layer in conventional UV LED structures. This causes severe internal light absorption and leads to a significant reduction of the light extraction efficiency.[8] Meanwhile, due to the extremely low thermally activated hole density and low hole mobility, poor hole transport in the p-type layers contributes to high operation voltage.[9] Therefore, both the light extraction efficiency and electrical efficiency face fundamental challenges for the conventional UV LED structures.

Recently, we demonstrated a tunnel-injected UV LED structure to address both the absorption and electrical loss issues.[10-13] We replaced the direct p-type contact using an interband tunneling contact by taking advantage of the polarization properties of III-Nitride material.[14-19] This minimizes internal light absorption caused by the p-GaN and p-type metal contact layers, and at the same time increases the hole injection efficiency.[10-13] Using this tunnel-injected UV LED structure, we have demonstrated efficient UV light emission at 325 nm with an on-wafer external quantum efficiency of 3.37%.[10,11,13] Simulations further showed that efficient interband tunneling could be achieved for high Al content AlGaN by using a compositionally graded tunnel junction structure.[12] It demonstrated the feasibility of achieving tunneling hole injection into deep UV LEDs.

The tunnel-injected UV LED structure enables n-type contacts for both bottom and top contact layers.[10-13,19-25] However, the fabrication of low resistance Ohmic contact to high Al composition AlGaN layers has been challenging. Even though vanadium or Titanium based metal contacts have been optimized to reduce contact resistance, low contact resistance below $1\times10^{-4}$ $\Omega$ cm$^2$ for high Al content (>75%) AlGaN has not been reported.[26-28] Meanwhile, these approaches require high temperature annealing and correspondingly metal spike into the material. This is not feasible for the top contact since metal diffusion into active region can lead to shunt leakage paths that degrade device performance. Our recent report showed that low contact resistance to ultra-wide bandgap $Al_{0.75}Ga_{0.25}N$ channel can be achieved through a reverse compositionally graded n++ AlGaN contact layer.[29] In this work, we apply the graded contact to our tunnel-injected UV LED structure, and demonstrate our first ultra-wide bandgap $Al_{0.75}Ga_{0.25}N$ tunnel junctions. Interband tunneling hole injection is confirmed from the electroluminescence measured at 257 nm.

Epitaxial stack of the tunnel-injected UV LED structure investigated in this work is shown in Fig. 1(a). The structure was grown by N$_2$ plasma assisted molecular beam epitaxy (MBE) on Si-doped metal-polar $Al_{0.72}Ga_{0.28}N$ template with a threading dislocation density of $3\times10^9$ cm$^{-2}$.[10-13] The template was grown on sapphire substrate using metal–organic chemical vapor deposition (MOCVD). The MBE growth was initiated with a 600 nm n+-$Al_{0.75}Ga_{0.25}N$ bottom contact layer with Si doping density of $1.8\times10^{19}$ cm$^{-3}$, followed by 50 nm n-$Al_{0.75}Ga_{0.25}N$ cladding layer ([Si]=$4\times10^{18}$ cm$^{-3}$), three periods of 2 nm $Al_{0.6}Ga_{0.4}N$/ 6 nm $Al_{0.75}Ga_{0.25}N$ quantum wells (QWs)/ barriers, 6 nm AlN electron blocking layer, 50 nm compositionally graded p-AlGaN layer, 4 nm $In_{0.2}Ga_{0.8}N$, 5 nm graded n++-AlGaN with Al content increasing from 62% to 75%, 200 nm n++ $Al_{0.75}Ga_{0.25}N$, and a 40 nm graded n++ AlGaN top contact layer. The p-AlGaN layer has a linear Al compositional grading from 95% to 65% to create a negative bulk polarization charge, which has been demonstrated to be useful in assisting acceptor activation and increasing hole density.[13,30] The 40 nm graded n++ AlGaN top contact layer has a similar Al compositional grading from 75% to 15%. The negative polarization charges behave like p-type doping in this layer, therefore, heavy Si doping to [Si]= $1\times10^{20}$ cm$^{-3}$ is used to compensate them.[29] This leads to a flat conduction band profile and effectively n-type doped layer.

Inductively coupled plasma reactive ion etching (ICP-RIE) with $BCl_3/Cl_2$ chemistry was used to reach n+-$Al_{0.75}Ga_{0.25}N$ bottom contact layer for device mesa isolation. V(20 nm)/Al(80 nm)/Ti(40 nm)/Au(100 nm) metal stack was deposited, and annealed at 860 °C for 3 min to form bottom contact.[27] Non-alloyed Al(30 nm)/Ni(30 nm)/Au(150 nm)/Ni(20 nm) metal stack was then deposited for top contact. The top contact was designed to have partial metal coverage on the mesa area. For the investigated 30 × 30 µm² devices, the metal contact covered 37% of the mesa region. This was followed by a low power ICP-RIE etch to remove the down-graded n++ AlGaN top contact layer within the device mesa region that has no metal contact to minimize internal light absorption. The final top contact schematic structure is shown in Fig. 1(b).

The forward biased energy band diagram under the top metal contact region is shown in Fig. 1(c). The graded contact layer enables smooth access to the heavily doped n-$Al_{0.75}Ga_{0.25}N$ layer for electrons.[29] The top n-AlGaN layer forms an interband tunneling contact to the p-AlGaN layer through a sharp band bending enabled by the thin InGaN layer.[10-13,16,19] Holes can therefore be injected by reverse biasing the tunnel junction structure. The reverse graded p-AlGaN layer provides flat valence band profile for hole injection into the active region, while it creates a high barrier to block the overflowing electrons, making this grading scheme beneficial for enhanced carrier injection efficiency.[13]

The non-alloyed top contacts and the alloyed bottom contacts are analyzed by transfer length measurement (TLM) and circular TLM (CTLM) methods, respectively. The top contact shows Ohmic behavior with an extracted specific contact resistance of $4.8\times10^{-5}$ Ω cm², which represents a combination of the resistance at metal/ AlGaN interface and the resistance of the reverse-graded contact layer. The sheet resistance extracted for the n++ $Al_{0.75}Ga_{0.25}N$ top current spreading layer is 1.1 kΩ/□. In comparison, the annealed bottom contact exhibits Schottky performance, which is consistent with recent investigations on metal/ AlGaN contacts with high Al content.[26] The sheet resistance of the bottom contact layer is estimated to be 2.3 to 2.5 kΩ/□ based on the CTLM measurement. The estimated sheet resistance drops with increased current flow as shown in Fig. 2(d). The deviation is attributed to the influence of the Schottky contact resistance on the adopted CTLM model. The performance contrast between the non-alloyed Ohmic top contact and the alloyed Schottky bottom contact demonstrates the benefit of using a down-graded AlGaN contact layer for contacts to high Al content n-AlGaN layers.

Two tunnel-injected UV LEDs were grown and fabricated under similar conditions as discussed above, with the only difference being the Mg doping concentration in the p-AlGaN layers. Mg doping was varied by controlling Mg flux during growth, and was calibrated using secondary ion mass spectrometry (SIMS) measurement as shown in Fig. 3(b). An exponential increase in the Mg doping concentration was observed with increasing Mg cell temperature at the p-AlGaN growth temperature of 715 °C. Because of the high growth temperature, Mg cell was operated at an upper temperature limit to provide sufficient doping concentrations in the two samples, resulting in [Mg]=$6.2 \times 10^{18}$ cm$^{-3}$ and $1.5 \times 10^{19}$ cm$^{-3}$ in sample A and B, respectively.

The current-voltage characteristics for the $30 \times 30\,\mu m^2$ devices are shown in Fig. 3(a). The samples showed similar voltage drop at 20 A/cm$^2$ (10.4 V for sample A and 10.2 V for sample B). This extra voltage drop as compared to the Al$_{0.6}$Ga$_{0.4}$N quantum well bandgap ($E_{g\_QW}$=4.9 eV) is attributed to contributions from the tunnel junction layer, the electron blocking layer and the alloyed bottom contact. In contrast, significant difference in voltage drop was observed between sample A and B at high current levels. The forward voltage at 1 kA/cm$^2$ reduced from 30.9 V to 19.2 V as the Mg doping concentration is increased. This is attributed to an extended depletion in the p-AlGaN layer due to the low Mg concentration in sample A when the tunnel junction layer is reverse-biased.[13] Since the tunneling probability drops exponentially with increasing barrier width, a high voltage drop across the tunnel junction layer is required to obtain efficient interband tunneling.

On-wafer electroluminescence (EL) measurement was carried out under continuous-wave operation to confirm interband tunneling hole injection. The emission spectrum was obtained using a calibrated Ocean Optics spectrometer by collecting light from the top surface of the $30 \times 30\,\mu m^2$ devices.[11] The EL spectrum of sample B with [Mg]=$1.5\times10^{19}$ cm$^{-3}$ is shown in Fig. 4. Single peak light emission at 257 nm was obtained. The microscope image shows uniform light emission from the whole device area even though the metal contact covers small part of the mesa area. The devices exhibited low efficiency, with a measured peak external quantum efficiency of 0.035%. This is the first demonstration of interband tunnel injected planar UV-C LED emitting at 257 nm using an AlGaN/ InGaN tunnel junction for hole injection.

We attribute the relatively low efficiency of the LEDs (compared to state-of-art UV-C LEDs[31], and longer wavelength tunnel injected LEDs[10-13]) to three reasons. Firstly, the active region growth has not been optimized, and the internal quantum efficiency may be low. Secondly, hole transport through the AlGaN/ InGaN tunnel junction and the p-type layers could be limiting LED performance. A high density of background compensating defects could cause low p-type density and poor hole injection.[13] Further optimization of the growth conditions and epitaxial structure could enable improvements in the efficiency.

In summary, interband tunneling hole injection through a polarization engineered Al$_{0.75}$Ga$_{0.25}$N/ In$_{0.25}$Ga$_{0.75}$N tunnel junction was demonstrated in a tunnel-injected UV-C LED structure. A compositionally graded top contact layer was used to form low resistance ($\rho_c$=$4.8\times10^{-5}$ $\Omega$ cm$^2$) non-alloyed Ohmic contact. We also observed large reduction in the forward operation voltage by increasing Mg doping concentration. This is attributed to enhanced interband tunneling because of the decrease in the p-AlGaN depletion barrier. Tunneling hole injection into p-AlGaN layers with $E_g$>5.2 eV enabled light emission at 257 nm with an on-wafer EQE = 0.035%. This work demonstrates the feasibility of tunneling hole injection into deep UV LEDs, and provides a structural design towards high power UV emitters.

Acknowledgement:

We acknowledge funding from the National Science Foundation (ECCS-1408416 and PFI AIR-TT 1640700). Sandia National Laboratories is a multi- mission laboratory managed and operated by Sandia

Figure captions:

Fig. 1 (a) Epitaxial stack of the tunnel-injected UV-C LED. (b) Schematic structure of the top contact layer. The down-graded n++ AlGaN top contact layer remains under the top contact metal. (c) Forward biased energy band diagram under the top metal contact region.

Fig. 2 (a) TLM measurement result of the top contact layer with non-alloyed metal contact, and (b) the change of extracted resistance with TLM pattern spacing. (c) Circular TLM measurement result of the bottom contact layer with annealed metal contact, and (d) the extracted sheet resistance at different current levels. The spacing for the CTLM pattern increases from 10 μm to 80 μm. The insets to (b) and (d) are the schematic structures for the TLM and CTLM measurements, respectively.

Fig. 3 (a) Current-voltage characteristics of the tunnel-injected UV-C LEDs with different Mg doping concentrations ([Mg]=$6.2\times10^{18}$ cm$^{-3}$ and $1.5\times10^{19}$ cm$^{-3}$ in sample A and B, respectively). The inset shows the current in log scale. (b) Mg concentration in $Al_{0.75}Ga_{0.25}N$ as a function of Mg cell temperature at the growth temperature of 715 °C as determined from SIMS measurement.

Fig. 4 (a) EL spectrums, (b) output power and EQE of the $30 \times 30$ μm$^2$ tunnel-injected UV-C LED measured under continuous-wave operation. Single peak emission at 257 nm was obtained. The results were measured on-wafer from the top surface. Inset of (b) shows a microscope image of a device with partial top metal coverage operated at 556 A/cm$^2$.

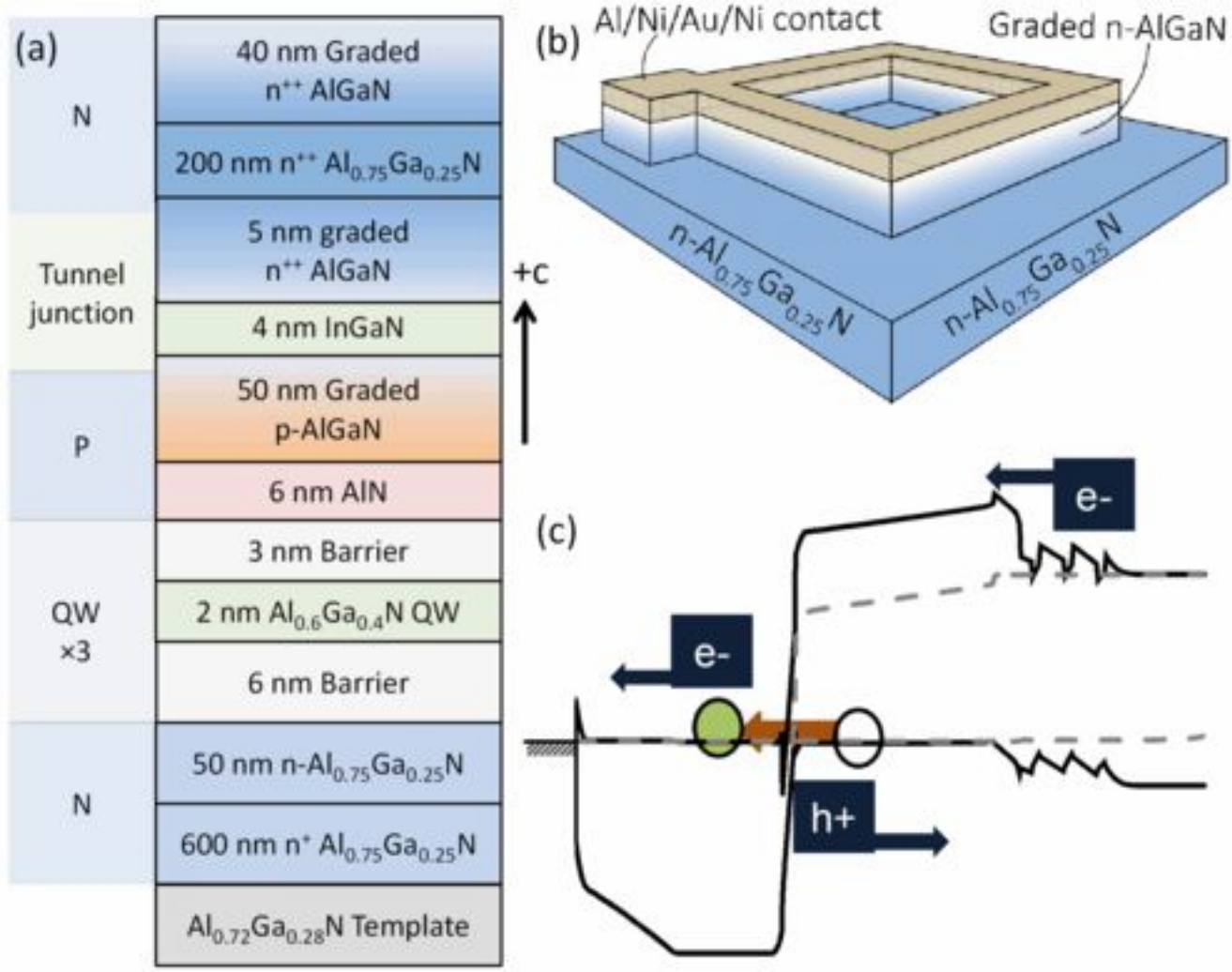

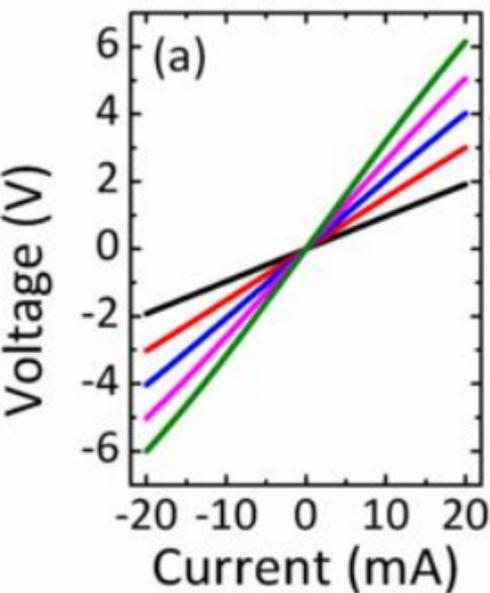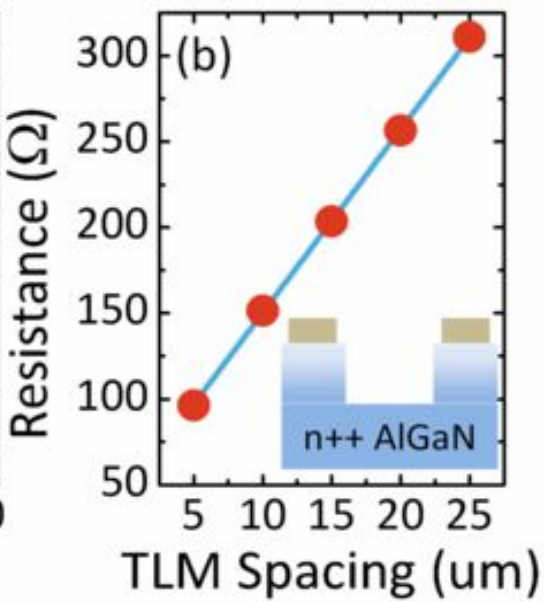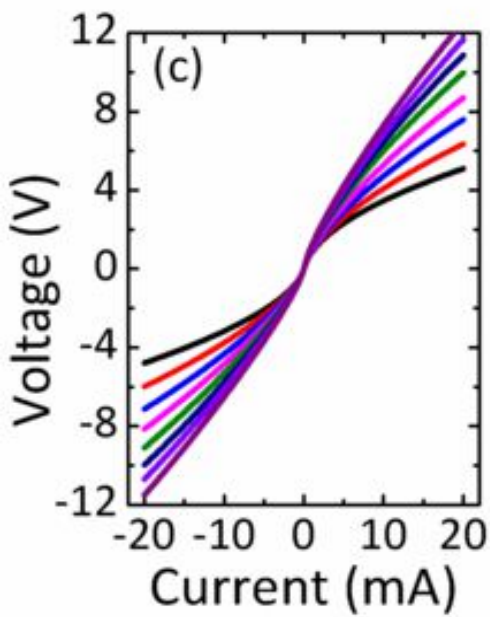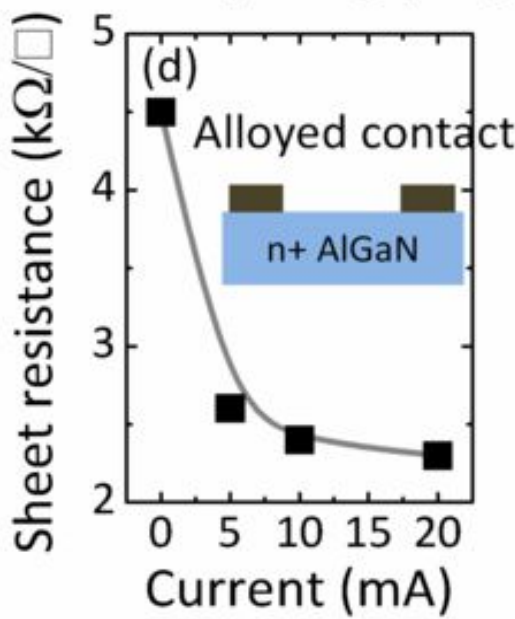

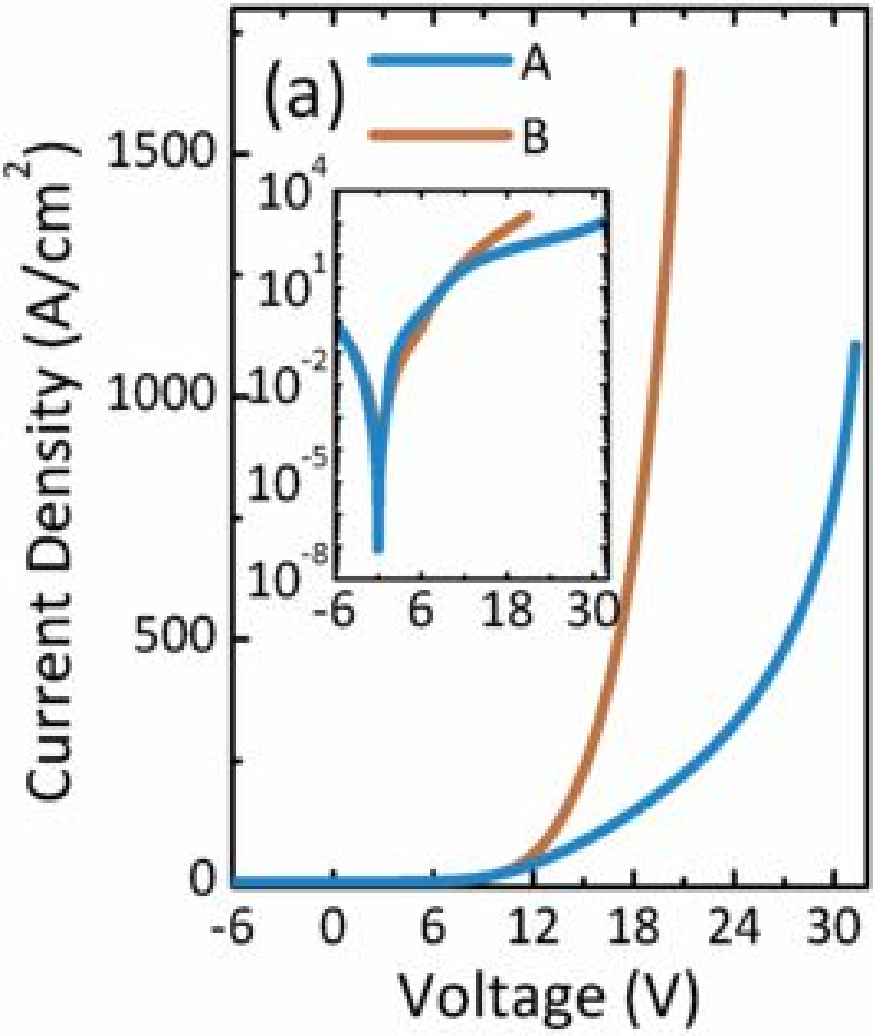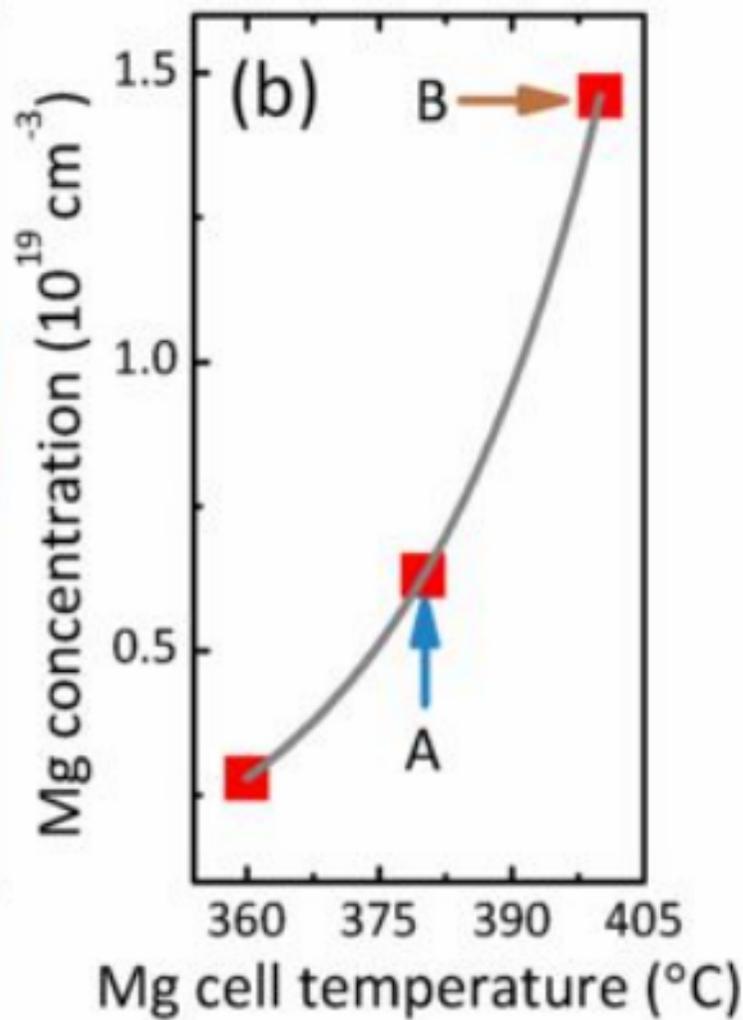

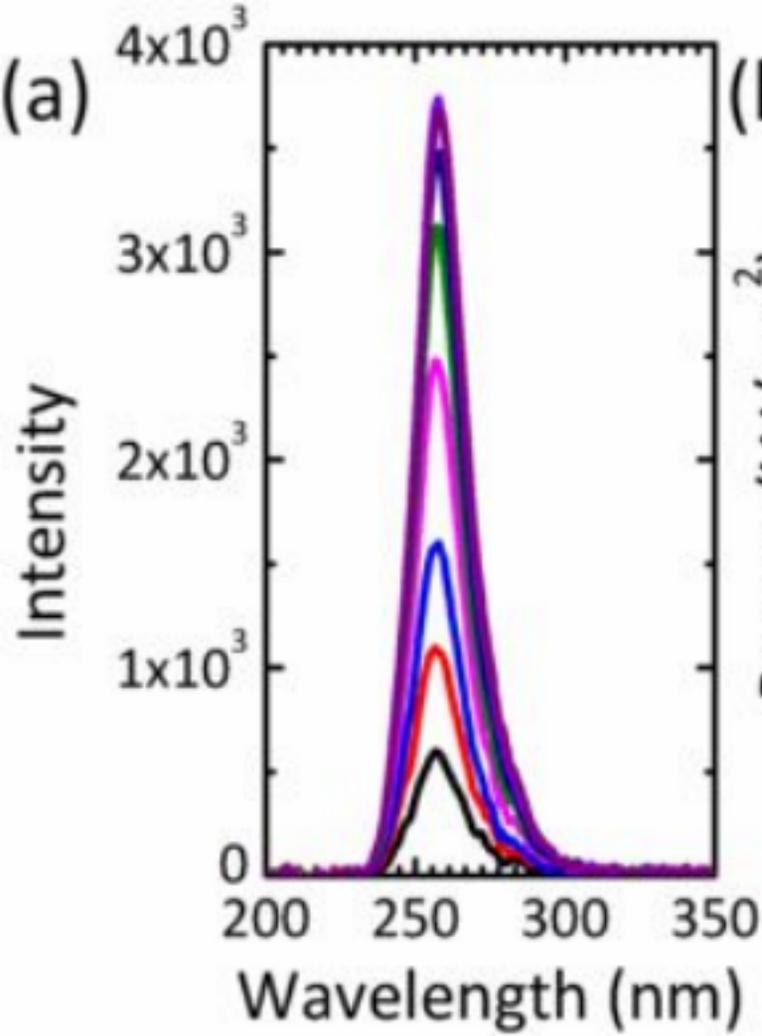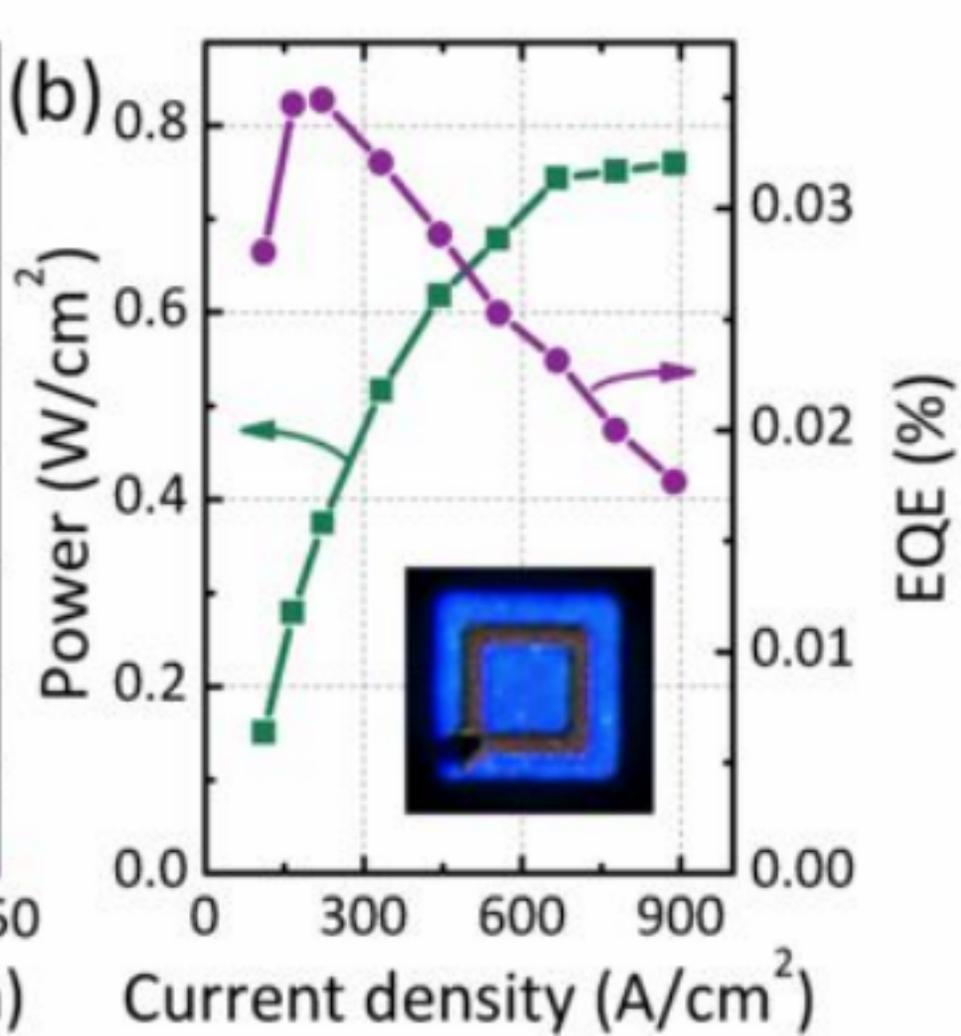